# TV White Space and LTE Network Optimization towards Energy Efficiency in Suburban and Rural Scenarios

Rodney Martinez Alonso, *Student Member, IEEE*, David Plets, *Member, IEEE*, Margot Deruyck, Luc Martens, *Senior Member, IEEE*, Glauco Guillen Nieto and Wout Joseph, *Senior Member, IEEE*

*Abstract*— The radio spectrum is a limited resource. Demand for wireless communication services is increasing exponentially, stressing the availability of radio spectrum to accommodate new services. TV White Space (TVWS) technologies allow a dynamic usage of the spectrum. These technologies provide wireless connectivity, in the channels of the Very High Frequency (VHF) and Ultra High Frequency (UHF) television broadcasting bands. In this paper, we investigate and compare the coverage range, network capacity, and network energy efficiency for TVWS technologies and LTE. We consider Ghent, Belgium and Boyeros, Havana, Cuba to evaluate a realistic outdoor suburban and rural area, respectively. The comparison shows that TVWS networks have an energy efficiency 9-12 times higher than LTE networks.

## I. Introduction

WIRELESS communication services are mainly provided under a fixed spectrum allocation. This spectrum allocation process is highly inefficient, leading to significant spectrum underutilization [1]. A radio spectrum usage survey in Virginia, United States, from 30 MHz to 3 GHz, revealed that less than 20% is in use at any location and at any given time [1]. A survey in Brno, Czech Republic and Paris, France indicated even a lower spectrum usage in the range from 400 MHz to 3 GHz [2]. A study to account the percentage of TV White Spaces (TVWS) in 11 European countries revealed that 56% of UHF spectrum is not in use at any location and at any given time [3]. The estimated percentage of unused UHF spectrum in Belgium is 69% [3]. Although assigned by the local regulatory domain, only 44% of VHF and UHF spectrum is in use in Havana City and it will decrease to 12% after analog broadcasting switch-off [4].

Manuscript submitted for review on April 6, 2017. R. Martinez Alonso is supported by LACETEL, and a doctoral grant from the Special Research Fund (BOF) of Ghent University, Belgium. M. Deruyck is a Post-Doctoral Fellow of the FWO-V (Research Foundation – Flanders, Belgium).

D. Plets, M. Deruyck, L. Martens and W. Joseph are with the INTEC Ghent University, Technologiepark-Zwijnaarde 15, 9052 Gent, Belgium (e-mail:{david.plets,wout.joseph,margot.deruyck,luc1.martens}@ugent.be).

R. Martinez Alonso, is with LACETEL and INTEC Ghent University, Technologiepark-Zwijnaarde 15, 9052 Gent, Belgium (e-mail: rodney.martinezalonso@ ugent.be).

G. Guillen Nieto is with the LACETEL, 34515 Rancho Boyeros Ave., Boyeros, 19200, Havana, Cuba (e-mail: glauco@enet.cu).

TVWS technologies dynamically allocate the required spectrum. The spectrum allocation is performed by means of cognitive radios with local spectrum sensing techniques and/or a geo-location database [5], [6]. Two main TVWS standards have been established based on the new dynamic spectrum-sharing paradigm: IEEE 802.22 (latest update IEEE 802.22b) and IEEE 802.11af [7], [8], [9]. IEEE 802.22 was the first complete cognitive radio standard, including spectrum sensing techniques and geo-location capability with the provision to access a database that stores, by geographic location, the permissible frequencies and operating parameters [10]. An amendment in IEEE 802.11af enables geolocation database access to TVWS. The location algorithm allows the implementation of a closed-loop database. This database provides to Base Stations (BS) the white spaces availability, but also receives feedback from the geo-location of all network devices, their frequencies and emission footprints. By accessing and using this information, it is possible to coordinate and to make intelligent decisions about the most effective way to utilize the available spectrum [9], [11].

Several trials have been conducted worldwide to evaluate TVWS technologies [12]. In a trial with IEEE 802.22, a Bit Error Rate (BER) of $10^{-6}$ was reported at a distance of 6.3 km (one site measurement), for 3/4 64-QAM with an Equivalent Isotropic Radiated Power (EIRP) of 34.6 dBm, BS antenna height 20 m, receiver antenna height of 12 m and receiver antenna gain of 7.65 dBi [13]. A field trial for Line-Of-Sight (LOS) studied the bitrate versus coverage of a TVWS prototype in four outdoor measurement sites, considering different link margin and modulation schemes [14].

In [15], the authors analyzed the coverage for 802.11af BSs in a generic scenario, for different interference conditions and BS antenna heights. A bitrate performance comparison of TVWS technology and WiFi is presented in [16], considering the effect of interference and medium access congestion for Carrier Sense Multiple Access with Collision Avoidance (CSMA/CA) mode in IEEE 802.11af. Also [17] presented a throughput study of IEEE 802.11af for a rural area, considering the population density as reference. Several studies have investigated the white space channel availability for different interference considerations, protection margins and occupancy thresholds [3], [18], [19], [20].

The power consumption models and energy efficiency for Long-Term Evolution (LTE) networks have been widely studied (i.e. [21], [22], [23], [24], [25]). For TVWS these parameters have not been properly investigated. A power consumption measurement for two different TVWS hardware is reported in [26], but neither power consumption model, energy efficiency nor network optimization is investigated.

In this paper we compare the coverage, performance and energy efficiency of TVWS technologies and LTE, in a suburban and a rural scenario, for the first time according to the authors' knowledge. A network optimization towards reduced power consumption is performed. We consider realistic user and traffic densities provided by local network operators. A novel power consumption model for TVWS technologies is proposed.

The outline of this paper is as follows. In Section II we describe the suburban and rural scenarios, the technology link budgets, the power consumption models, the energy efficiency metric, and the optimization algorithm. In Section III, we present the network simulation results for the proposed scenarios. Conclusions are presented in Section IV.

## II. METHOD

First, we define the characteristics of each scenario and the link budget for each technology. The maximum coverage range and minimum required number of BS are estimated. Finally, an optimized network design towards minimum power consumption for TVWS and LTE is performed.

### A. Evaluation Scenarios

We consider Ghent City, Belgium and Boyeros municipality outskirts in Havana, Cuba for the evaluation of a realistic suburban and rural area, respectively. Fig. 1a shows the target area (68 km$^2$) that needs to be covered in Ghent City. Fig. 1b shows the target area (169 km$^2$) that needs to be covered in Boyeros municipality. This area also includes some small towns at the outskirts of Havana City with dispersed population.

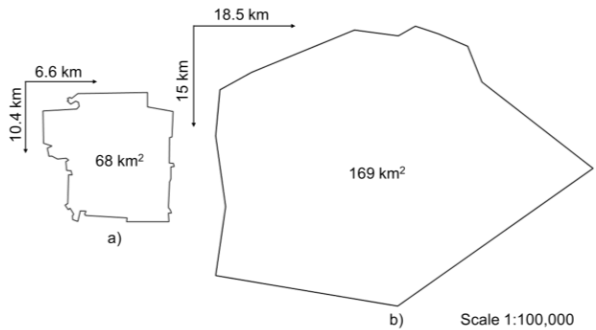

Fig. 1. The area to be covered, a) Ghent City (Suburban), b) Boyeros municipality – Havana outskirts (Rural).

We consider a wireless network setup based on a fixed outdoor over-roof antenna configuration. The end-point connection at the user's home is provided by a transceiver to an Ethernet or WiFi network. Fig. 2 shows the initial considered configuration.

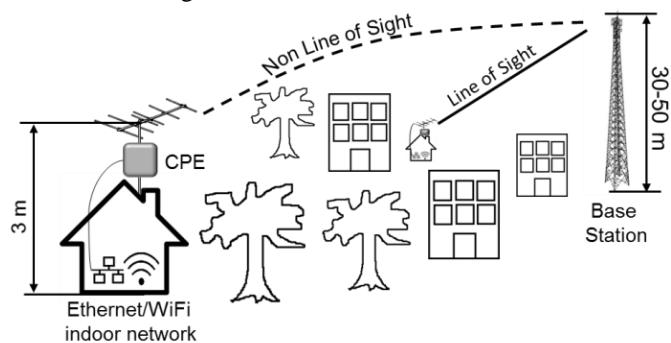

Fig. 2. Configuration for evaluation of TVWS and LTE technologies.

The wireless network design is based on a requirement of 90% of the locations covered at the edge of the coverage area during 99% of the time. The network must be able to handle up to 224 simultaneous connections in the suburban area [27] and 135 simultaneous connections in the rural area. These user densities are based on real statistical data from the local service providers for the busiest network time. Some users require 64 kbps (voice) and others 1 Mbps (data) [27]. The bitrate per user is pseudo-randomly and probabilistically distributed (approximately 91% data users and 9% voice users).

In both suburban and rural scenarios, the antenna configuration is Single-Input Single-Output (SISO) with an omnidirectional radiation pattern. Therefore, the BS coverage area is represented by a circle with center in the BS location coordinates. LTE, IEEE 802.22b and IEEE 802.11af provide support for Multi-Input Multi-Output (MIMO) up to 4x4, although it is not widely industrialized yet [29], [8], [9]. The highest Modulation and Coding Schemes (MCS) for IEEE 802.22b (256-QAM schemes) are not implemented on commercial available hardware either. The influence of a MIMO 4x4 configuration will be evaluated for both scenarios (considering future hardware availability).

### B. Link Budgets and Propagation Models

To estimate the range of each BS, the maximum allowable path loss PL$_{max}$ [dB] for an acceptable BER performance has to be determined [21], [30]. To this aim, a link budget is defined according to the technology specifications and the scenario characteristics [30].

Table I lists the link budget parameters for each technology, in both scenarios. Main link budget differences from one scenario to another one arise in the parameters related to the propagation environment (i.e. shadowing and fading margin) and the regulatory domain (i.e. bandwidth, frequency). These parameters are retrieved from specifications provided in the standards [7], [8], [11], [29] as well as values from manufacturers [31] or published research [32], [33], [34], [10].

TABLE I
LINK BUDGET PARAMETERS (SUBURBAN/RURAL AREA)

| Parameter | 802.22 | 802.22b | 802.11af | LTE | Unit |
|---|---|---|---|---|---|
| Radiated Power | 36 | 36 | 36 | 36 | dBm |
| Frequency | 602/605 | 602/605 | 602/605 | 821 | MHz |
| Bandwidth | 8/6 | 8/6 | 8/6 | 10/5 | MHz |
| Total Subcarriers | 2048 | 1024 | 144 | 1024/512 | - |
| Used Subcarriers | 1680 | 832 | 114 | 601/301 | - |
| Frequency Sampling Factor | 1.142 | 0.9325 | 1.142 | 1.536 | - |
| BS Antenna Height | 50/30 | 50/30 | 50/30 | 50/30 | m |
| Cell Interference Margin | 0 | 0 | 0 | 2 | dB |
| MIMO Gain | - | 12 | 12 | 12 | dB |
| Receiver Antenna Height | 3 | 3 | 3 | 3 | m |
| Receiver Antenna Gain | 11.5 | 11.5 | 11.5 | 11.5 | dB |
| Receiver Feeder Losses | 0.04 | 0.04 | 0.04 | 0.04 | dB |
| Noise Figure | 4 | 4 | 4 | 7 | dB |
| Shadow Margin | 7.91/5.5 | 7.91/5.5 | 7.91/5.5 | 7.91/5.5 | dB |
| Fade Margin | 7.37/4 | 7.37/4 | 7.37/4 | 7.37/4 | dB |
| Receiver Signal to Noise Ratio (SNR) | 4.3<br>10.2<br>12.4<br>18.3<br>19.7 | 4.3<br>10.2<br>12.4<br>18.3<br>19.7 | 3.8<br>8.0<br>15.1<br>25.2<br>30.4 | 3.0<br>10.5<br>14.0<br>22.0<br>29.4 | dB |
| | | 26.9<br>28.2 | | | |
| Bitrate | @ 8 MHz<br>6.0<br>12.0<br>16.1<br>24.1<br>27.2 | @ 8 MHz<br>6.0<br>12.0<br>16.1<br>24.1<br>27.2<br>32.2<br>42.3 | @ 8 MHz<br>2.4<br>7.2<br>14.4<br>24.0<br>32.0 | @ 10 MHz<br>4.32<br>6.3<br>16.8<br>25.2<br>38.7 | Mbps |
| | @ 6 MHz<br>4.5<br>9.0<br>12.1<br>18.1<br>20.4 | @ 6 MHz<br>4.5<br>9.0<br>12.1<br>18.1<br>20.4<br>24.2<br>31.7 | @ 6 MHz<br>1.8<br>5.4<br>10.8<br>18.0<br>24.0 | @ 5 MHz<br>4.2<br>5.7<br>8.5<br>11.3<br>16.9 | |

The cell interference margin for TVWS technologies is 0 dB. We consider TVWS operation in non-interfering channels only (25 non-interfering TVWS channels available in Ghent [35]) and the strictest spectrum-sensing modes defined in the standards. The channel allocation is based on the detection of the wireless beacon (IEEE 802.22.1). It means that a channel will be considered occupied if a wireless beacon frame, with a level equal or higher than -116 dBm is detected. For a complementary protection from/to the primary TV broadcasting services, we assume a geo-location database with a similar constraint.

Even under the considered constraints, the coexistence of joint IEEE 802.22 and IEEE 802.11af operating in the same region is not solved yet. In such condition, the probability of IEEE 802.22 users to get access to the spectrum resources is higher [36]. For a fair comparison, we assume that a single TVWS technology (IEEE 802.22 or IEEE 802.11af) is deployed in the target area at the same time. State-of-the-art TVWS receivers have a noise figure from 3 to 4 dB [31]. To evaluate the worst case scenario, all calculations are based on a noise figure of 4 dB. For LTE, the macrocell propagation model proposed by ETSI (European Telecommunications Standards Institute) considers a Customer-Premises Equipment (CPE) noise figure of 9 dB [37]. The noise figure of current fixed outdoor LTE receivers (in the frequency of interest) varies from 4 dB to 8 dB. We assume a noise figure of 7 dB for LTE fixed outdoor CPE (see Table I).

In suburban Ghent, we consider a shadow margin of 7.91 dB, for 90% of locations covered at a certain distance from the transmitter [38] and a fading margin of 7.37 dB for 99% availability [34]. Boyeros municipality in Havana outskirts, is a rural area with a low foliage density. To achieve the same coverage and availability percentages, we consider a shadow margin of 5.5 dB [39] and a fading margin of 4 dB [34]. Note that over-roof reception network planning does not require accounting for building penetration losses [28].

Fig. 3 shows the path loss models considered for the BS range calculation. An experimental suburban one-slope path loss model for Ghent City is used [38]; in the rural area, the BS range calculations are based on the Okumura-Hata path loss model for rural environments with non-dense foliage [40], since we do not dispose of an experimental model for this area.

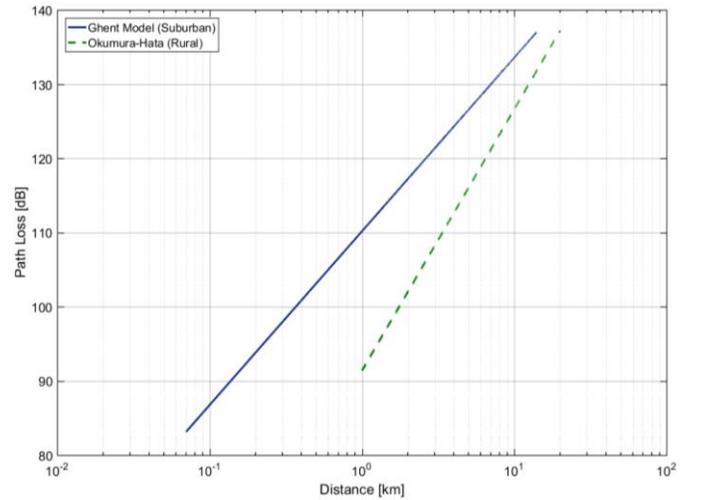

Fig. 3. Ghent (Suburban) and Okumura Hata (Rural) path loss models.

### C. Minimum required number of Base Stations

The minimum required number of BS depends on both the area to be covered and the served traffic. To cover a target area $A_T$ (km$^2$), the minimum required number of BS ($N_{BSa}$) can be defined as a function of the maximal BS coverage range

$R$ (km), with $\lceil \cdot \rceil$ the ceil function:

$$N_{BSa} = \left\lceil \frac{A_T}{\pi \cdot R^2} \right\rceil \quad (1)$$

Taking into consideration the total traffic requirement $T$ (Mbps) within the area $A_T$ (i.e. the sum of the individual traffic requirements of all simultaneous users), the minimum required number of BS ($N_{BSt}$) as a function of the bitrate served by a single BS, $B_{BS}$ (Mbps), can be defined by the following equation:

$$N_{BSt} = \left\lceil \frac{T}{B_{BS}} \right\rceil \quad (2)$$

The actually required number of BS ($N_{BS\_total}$) depends also on the target area topology, the distribution of possible BS locations and the network optimization algorithm. Equation 3 provides a minimum to the required number of BS.

$$N_{BS\_total} \geq \max(N_{BSa}, N_{BSt}) \quad (3)$$

### D. Energy Efficiency Metric

A metric to account for the energy efficiency of a single BS is defined in [22]. An extension of this metric, to account for the energy efficiency of the whole network configuration is defined in [41]. The average network energy efficiency $EE_n$ (km$^2$·Mbps/W) for $t$ different user distributions and a certain coverage percentage of users $c_i$ can be defined as follows:

$$EE_n = \frac{1}{t} \cdot \sum_{i=1}^{t} \frac{c_i \cdot A_T \cdot U \cdot \sum_{j=1}^{m} B_{ij}}{\sum_{j=1}^{m} P_{BS_{ij}}} \quad (4)$$

where $A_T$ is target area, $U$ is the number of users in the target area, $B_{ij}$ represents the total bitrate provided by BS $j$ to the users population $i$, $m$ is the total number of BS and $P_{BSij}$ represents the power consumption of the $BS_{ij}$.

The power consumption of LTE BS has been studied in [25], [22], [24]. To account for the power consumption of each LTE BS we consider the model proposed in [21]. This model takes into account the radiated power, the amplifier efficiency and the radiation system efficiency. Fig. 4a shows the power consuming components of an LTE BS. We assume an optical backhaul power consumption of 32 W for LTE [42].

In Fig. 4b we propose a power consumption model for the TVWS BS. This model comprises three power-consuming components: the Radio Unit (RU), the Power Supply (Power over Ethernet (PoE)) and the Optical Backhaul.

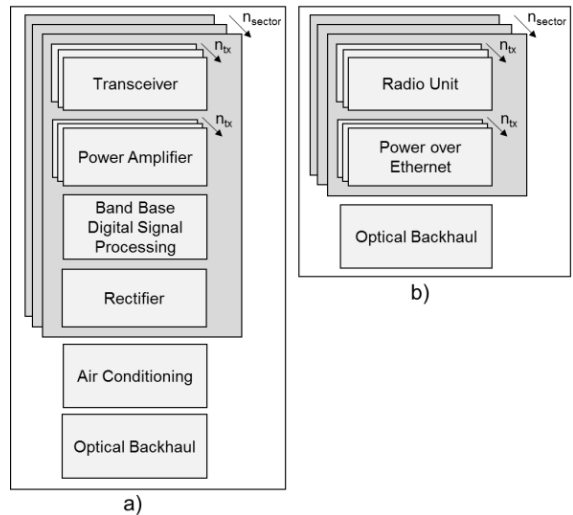

Fig. 4. Block diagram of power-consuming components a) LTE b) TVWS.

The total power consumption of a TVWS BS can be calculated as follows:

$$P_{BS} = P_{bh} + P_{idle} + n_{st} \cdot n_{tx} \cdot \alpha \cdot \left( \frac{P_r}{\eta_{ru}} + P_{PoE} \right) \quad (5)$$

We consider the power consumption of the optical backhaul $P_{bh}$ to be constant and independent from the number of sectors and transmitters. The power consumption of the RU can be correlated with the radiated power $P_r$ (for $n_{st}$ sectors and $n_{tx}$ transmitters) by means of the RU efficiency factor $\eta_{ru}$. The PoE power consumption ($P_{PoE}$) and RU power consumption varies with the traffic load factor $\alpha$. We will consider $\alpha = 1$ to investigate the maximal power consumption of the BS (worst-case scenario). The RU also has an idle power consumption $P_{idle}$ (6W) for $P_r = 0$ W and $\alpha = 0$. Table II lists the power consumption values for a TVWS BS.

TABLE II
TVWS POWER MODEL PARAMETERS

| Parameter | Value | Unit |
|---|---|---|
| $P_{bh}$ [42] | 32 | W |
| $P_{PoE}$ [43] | 4 | W |
| $P_{idle}$ [31] | 6 | W |
| $\eta_{ru}$ | 0.182 | - |

The RU efficiency is calculated based on the data provided in [19]. For a single transmitter and sector with $P_r = 4$ W and $\alpha = 1$ the power consumption of a TVWS BS is as low as 64 W. Digital Signal Processing (DSP) on TVWS BSs is generally implemented by a dedicated chipset (i.e. [31]). This is highly energy efficient [44] but, as a drawback, BSs are not upgradeable to support future standards or re-scale.

### E. Optimization algorithm

The network planning is performed by GRAND (Green Radio and Access Network Design) optimization algorithm described in [27]. First, the network traffic is generated for 40 simulations (40 different spatial user distributions and user bitrate distributions). Fig. 5 shows the optimization algorithm towards minimization of the network power consumption (for one simulation).

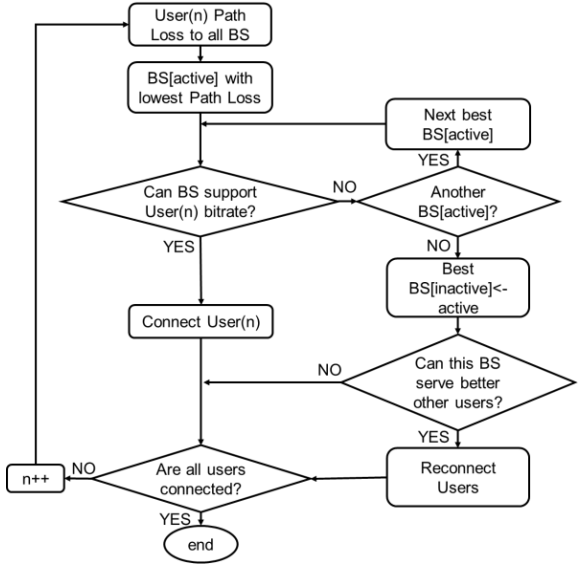

Fig. 5. Network optimization algorithm *[27]*.

For each simulation, the software calculates the path loss between a user and all possible BSs. The current user will be connected to the BS corresponding with the lowest path loss (for optimal power consumption); if this BS is already active and still can support the bitrate demanded by the user. In case this BS can not support the current user the algorithm seeks for the next already active BS with lowest path loss. In case no active BS can be found, the algorithm will activate the most appropriate BS (lowest path loss) from the inactive ones (Fig. 5). When a new BS is activated, the algorithm checks if users already connected can be switched in order to balance the network load. A certain user is only switched to another BS if the pass loss to the new BS is lower. The described algorithm is repeated until all users are evaluated. [27]

The progressive average for all simulations is calculated to validate a proper estimation of the percentage of users covered.

## III. RESULTS

This section presents the results of the network simulations and optimizations for the considered scenarios.

### A. Maximum coverage for one Base Station

Fig. 6 shows the BS bitrate versus coverage range for the suburban and rural area. For the maximum EIRP, IEEE 802.22b BS has a higher coverage range than LTE, IEEE 802.22 and IEEE 802.11af. The maximum coverage range for IEEE 802.22b is equal to 7.0 km in the suburban scenario and 17.6 km in the rural scenario (MCS 1/2 QPSK). The LTE BS has the lowest coverage range: 3.2 km and 12.1 km (1/2 QPSK), in the suburban and rural area, respectively. The lower coverage is due to a 3 dB higher noise factor compared with TVWS technologies and 2 to 4 dB higher required SNR than IEEE 802.22b. The latest version IEEE 802.22b achieves a 5 to 8% higher coverage range than IEEE 802.22 due to an improvement in the ratio of OFDM used from total subcarriers and a better sampling frequency factor. In comparison, IEEE 802.11af achieves a 15 to 30% lower coverage because it requires a higher SNR (an average 6 dB higher).

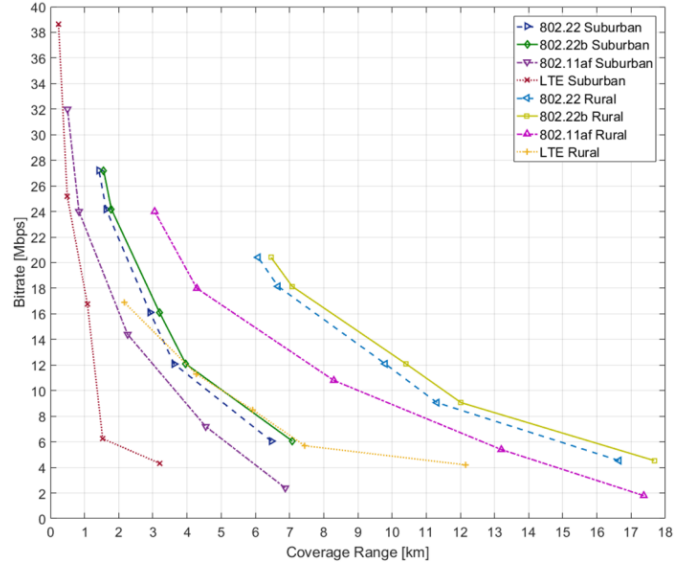

Fig. 6. BS comparison of bitrate versus coverage for the maximum EIRP.

### B. Network planning and optimization

First, we simulate the network for the minimum required number of BS (see Equation 3). Fig. 7 and Fig. 8 show the minimum required number of BS in the suburban and rural scenario, respectively. These graphs represent the trade-off between coverage and capacity. For a higher bitrate (i.e. higher required SNR), the number of BS to satisfy a certain traffic ($N_{BSt}$) decreases. A higher SNR has as consequence, a reduction in coverage and the number of BS to cover a certain area ($N_{BSa}$) increases. IEEE 802.22(b) has the lowest required number of BS. This is because it has a better coverage per provided bitrate unit in both scenarios (see Fig. 6).

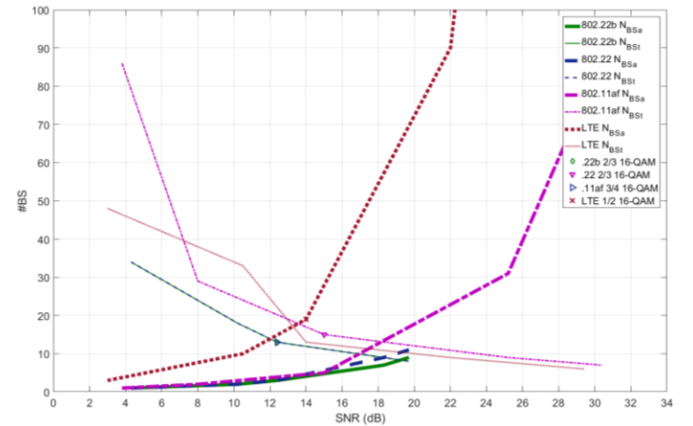

Fig. 7. Minimum required number of BS vs required SNR in the **suburban** scenario.

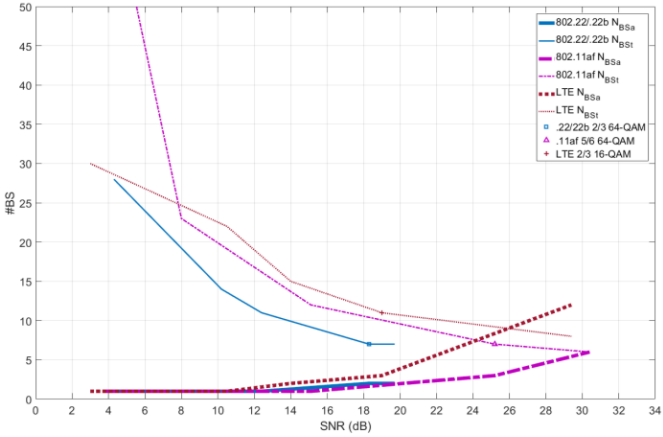

Fig. 8. Minimum required number of BS vs required SNR in the **rural** scenario.

The minimum required number of BS locations in the rural area is lower for all the standards, due to a better propagation environment and lower traffic density.

In the suburban scenario, the lowest required number of BSs and highest percentage of users covered are achieved for the MCS 2/3 16-QAM for IEEE 802.22 and IEEE 802.22b, 3/4 16-QAM for IEEE 802.11af and 1/2 16-QAM for LTE (see markers in Fig. 7). In the rural scenario, the best trade-off is achieved for the MCS 2/3 64-QAM for IEEE 802.22 and IEEE 802.22b, 5/6 64-QAM for IEEE 802.11af and 2/3 16-QAM for LTE (see markers in Fig. 8). Note that in the suburban scenario the MCS 2/3 16-QAM (IEEE 802.22(b)) does not has the lowest **max** ($N_{BSa}$, $N_{BSt}$). Nevertheless, due to the deviation caused by the area geometry and BS location distribution, for MCS with a similar minimum number of required BS, the area constraint ($N_{BSa}$) prevails.

Equation 3 provides a minimum $N_{BS\_total}$ been required a higher number of BS. This is due to the area geometry and BS location influence. Therefore, the number of BS is increased until we reach a mean coverage higher than 95%.

Fig. 9 shows the network coverage map for each technology in both scenarios. In the suburban scenario, the number of considered BS locations is 20 for IEEE 802.22 and 802.22b, 21 for IEEE 802.11af and 36 for LTE. In the rural scenario, the number of considered BS locations is 10 for all TVWS technologies and 13 for LTE. Although the target area in the rural scenario is more than two times larger than the suburban scenario, the number of BS locations can be reduced around to half (keeping a similar percentage of coverage). This is because the rural environment has a lower path loss and lower traffic requirement.

The mean percentage of users covered is 95% for LTE and higher than 96% for TVWS technologies in the suburban scenario. In the rural scenario the mean percentage of users covered is higher than 96% for IEEE 802.11af and LTE, and higher than 99% for IEEE 802.22(b). The deviation of the mean value of the percentage of users covered is lower than 0.5% over the considered simulations.

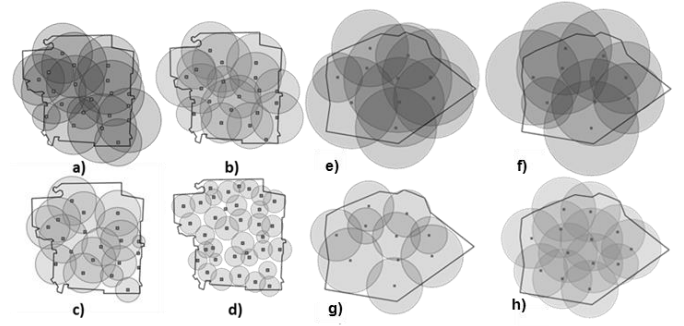

Fig. 9. Optimized networks (towards minimum power consumption) in Ghent area (suburban), for a) IEEE 802.22b, b) IEEE 802.22, c) IEEE 802.11af, d) LTE technology and Boyeros area (rural) for e) IEEE 802.22b, f) IEEE 802.22, g) IEEE 802.11af, h) LTE technology.

Fig. 10 shows the average network energy efficiency and its standard deviation for each technology. The best solution to cover the suburban area is IEEE 802.22b with an average network energy efficiency of 2996.8 km$^2$·Mbps/W. Note that the energy efficiency difference between IEEE 802.22b and IEEE 802.22 is lower than the standard deviation.

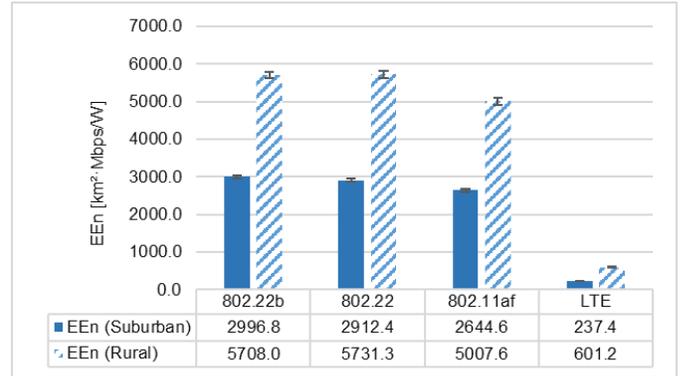

Fig. 10. Average network energy efficiency in the suburban and rural scenarios.

The LTE network has an energy efficiency more than 12 times lower. This is due to a lower coverage per provided bitrate unit (see Fig. 6) but also a higher network power consumption. The total power consumption for TVWS networks ranges from 1010 W (IEEE 802.22b) to 1044 W (IEEE 802.22). The full LTE network has an average power consumption of 13769 W.

The energy efficiency in the rural scenario is higher than the energy efficiency in the suburban scenario due to a better propagation environment and lower traffic density. This leads towards a lower number of BSs and lower network power consumption. The network power consumption for the TVWS networks ranges from 489 W (IEEE 802.22) to 521 W (IEEE 802.11af) and 4362 W for LTE.

The best solutions to cover the rural area are IEEE 802.22b and IEEE 802.22 (difference lower than the standard deviation). This is because the slightly difference in coverage per bitrate provided (see Fig. 6) is not enough to compensate the deviation caused by the BS location distribution and area geometry.

## C. Influence of MIMO

The diversity gain increases the coverage of each BS. As a consequence, the number of BSs can be reduced. In the suburban scenario with a MIMO 4x4 configuration the number of BSs can be reduced to 11 for TVWS technologies and 15 for LTE. The best trade-off between area covered and capacity is achieved for 3/4 64-QAM for IEEE 802.22b, 5/6 64-QAM for IEEE 802.11af and 1/2 16-QAM for LTE. In the rural scenario, the number of required BSs can be reduced to 5 for IEEE 802.22b, 8 for IEEE 802.11af and 10 for LTE.

Fig. 11 shows the average network energy efficiency for MIMO 4x4 compared with SISO configuration. In the suburban scenario, the energy efficiency of IEEE 802.22b with a 4x4 MIMO configuration is slightly increased by 4% while for LTE by 47% (compared with SISO). For IEEE 802.11af the energy efficiency decreases approximately 15% when comparing with SISO. In the suburban scenario, four transmitting antennas significantly increases the network energy efficiency for LTE but not for TVWS. For LTE BSs, the transmitters consume less than 10% of the BS total consumed power. The coverage increase realized by the MIMO diversity gain, together with the reduction of BS sites, overcompensates the increase in the transmitters' power consumption. For TVWS BS the power consumption of the transmitters represents around 40% of the total consumed power. The increase in the power consumption of the transmitters is not always compensated.

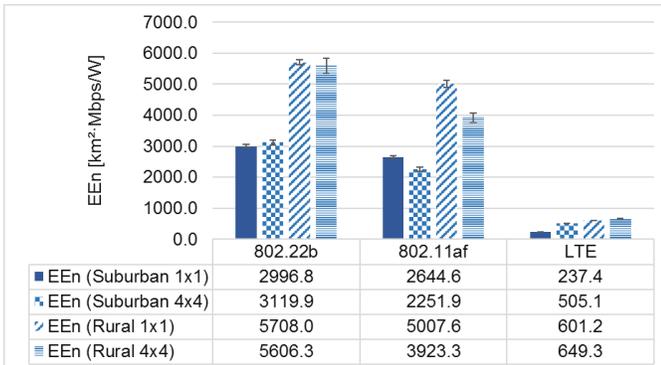

Fig. 11. Average network energy efficiency in the suburban and rural scenario. MIMO 4x4 versus SISO configuration.

For the rural scenario, the minimum required number of BS is always defined by the traffic constraint. Only 1 to 3 BS are required to cover the whole area however these can not support the traffic demand. The usage of four transmitters leads towards a higher power consumption not compensated by the coverage increase. As consequence, for the rural scenario the energy efficiency slightly increases by 7% for LTE, remaining approximately the same for IEEE 802.22b (the difference is less than the standard deviation), while decreases more than 21% for IEEE 802.11af. Nevertheless, all the technologies have a better performance in the rural scenario, prevailing the better propagation conditions.

## IV. CONCLUSION

By using novel network planning software, we investigated the coverage, capacity and energy efficiency of TVWS networks, optimized towards reducing its power consumption, in realistic suburban and rural scenarios. For this aim, we proposed a model to determine the energy consumption of TVWS networks. We also optimized and investigated an LTE network for a reference comparison. This comparison reveals that LTE has a lower energy efficiency in both suburban (approximately 12 times lower) and rural (approximately 9 times lower) scenario. IEEE 802.22b achieves the highest energy efficiency (12% higher than IEEE 802.11af).

For TVWS technologies, the use of a MIMO 4x4 configuration allows reducing the number of BS locations but does not significantly increase the energy efficiency in the considered scenarios.

Future research will consist of planning energy efficient Internet of Things (IoT) wireless networks in TVWS band. A huge density of devices will have to be considered and coverage, capacity and density will play a key role.

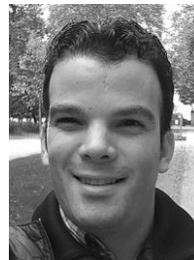

**Rodney Martinez Alonso** was born in 1987 in Havana, Cuba. In 2010, he obtained a B.Sc. degree in Telecommunications and Electronics Engineering and a M.Sc. degree on Digital Systems (2015) from the Higher Polytechnic Institute CUJAE, Havana, Cuba. Since 2010, he is a fellow researcher at LACETEL, R&D Telecommunications Institute. He also has collaborated on DTV engineering projects with HAIER and KONKA (China). Currently, he is a PhD student at WiCa group (Department of Information Technology – INTEC, Ghent University).

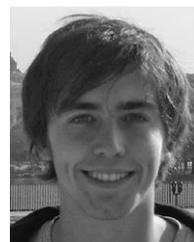

**David Plets** was born in 1983 in Belgium. In 2006, he obtained a Master in Electrotechnical Engineering, with ICT as main subject. Currently, he is a member of the WiCa group (Department of Information Technology – INTEC, Ghent University). In 2011, he obtained his PhD with a dissertation on the characterization and optimization of the coverage of wireless broadcast and WLAN networks. His current research interests include low-exposure wireless indoor network planning, cognitive networks, WiFi QoS optimization, and localization algorithms.

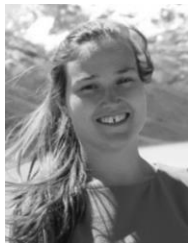
**Margot Deruyck** Margot Deruyck was born in Kortrijk, Belgium, on July 14, 1985. She received the M. Sc. degree in Computer Science Engineering and the Ph. D. degree from Ghent University, Ghent, Belgium, in 2009 and 2015, respectively. From September 2009 to January 2015, she was a Research Assistant with Ghent University - IMEC – WAVES (Wireless, Acoustics, Environment & Expert Systems – Department of Information Technology). Her scientific work is focused on green wireless access networks with minimal power consumption and minimal exposure from human beings. This work led to the Ph.D. degree. Since January 2015, she has been a Postdoctoral Researcher at the same institution where she continues her work in green wireless access network. Since October 2016, she is a Post-Doctoral Fellow of the FWO-V (Research Foundation - Flanders).

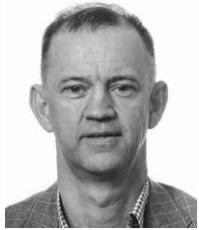
**Luc Martens** received the M.Sc. degree in electrical engineering from the Ghent University, Belgium in July 1986. From September 1986 to December 1990 he was a research assistant at the Department of Information Technology (INTEC) of the same university. During this period, his scientific work was focused on the physical aspects of hyperthermic cancer therapy. His research work dealt with electromagnetic and thermal modelling and with the development of measurement systems for that application. This work led to the Ph.D. degree in December 1990. Since 1991, he manages the wireless and cable research group at INTEC. This group is since 2004 part of the iMinds institute and since April 1993 he is Professor at Ghent University. His experience and current interests are in modelling and measurement of electromagnetic channels, of electromagnetic exposure e.g. around telecommunication networks and systems such as cellular base station antennas, and of energy consumption in wireless networks. He is author or co-author of more than 300 publications in the domain of electromagnetic channel predictions, dosimetry, exposure systems and health and wireless communications.

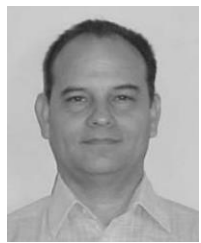
**Glauco Guillen Nieto** was born in 1961 in Havana, Cuba. In 1985, he obtained a B.Sc. degree in Radiocommunications and Broadcasting Engineering and a PhD. degree (1989) from the Electrotechnical Institute of Communication "A. S. Popov", Odessa, Ukraine. He is currently a senior researcher at LACETEL and elected member of the Cuban Academy of Science (2012-2018). From 2012, he is a "Series A" advisor of the local network operators. He also has collaborated in different projects with Finline Technology (Canada), DTVNEL (DTV National Engineering Lab - Beijing), HAIER and KONKA (China).

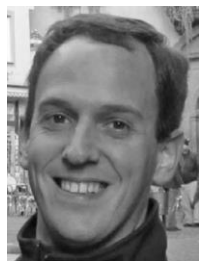
**Wout Joseph** was born in Ostend, Belgium on October 21, 1977. He received the M. Sc. degree in electrical engineering from Ghent University (Belgium), in July 2000. From September 2000 to March 2005 he was a research assistant at the Department of Information Technology (INTEC) of the same university. During this period, his scientific work was focused on electromagnetic exposure assessment. His research work dealt with measuring and modelling of electromagnetic fields around base stations for mobile communications related to the health effects of the exposure to electromagnetic radiation. This work led to a Ph. D. degree in March 2005. Since April 2005, he is postdoctoral researcher for iMinds-UGent/INTEC. From October 2007 to October 2013, he was a Post-Doctoral Fellow of the FWO-V (Research Foundation – Flanders). Since October 2009, he is professor in the domain of Experimental Characterization of wireless communication systems. His professional interests are electromagnetic field exposure assessment, in-body electromagnetic field modelling, electromagnetic medical applications, propagation for wireless communication systems, antennas and calibration. Furthermore, he specializes in wireless performance analysis and Quality of Experience.